\documentclass[onecolumn,nofootinbib,superscriptaddress,tightenlines,notitlepage,11pt]{revtex4-1}

\usepackage{array}
\usepackage{multirow}

\usepackage[ colorlinks = true,
             linkcolor = blue,
             urlcolor  = blue,
             citecolor = red,
             anchorcolor = green,
]{hyperref}

\usepackage{dsfont}
\usepackage{amsmath}
\usepackage{amssymb}
\usepackage{amsthm}
\usepackage{graphicx}
\usepackage{mathrsfs}
\usepackage{units}

\newcommand{\comment}[1]{}

\newcommand{\abs}[1]{\ensuremath{|#1|}}

\newcommand{\Abs}[1]{\ensuremath{\left|#1\right|}}

\newcommand{\norm}[2]{\ensuremath{|\!|#1|\!|_{#2}}}
\newcommand{\normbig}[2]{\ensuremath{\big|\!\big|#1\big|\!\big|_{#2}}}

\newcommand{\idx}[2]{{#1}_{\textnormal{\tiny #2}}}

\newcommand{\ket}[1]{| #1 \rangle}

\newcommand{\braket}[2]{\langle #1 | #2 \rangle}

\newcommand{\kron}{\otimes}
\newcommand{\eps}{\varepsilon}

\newcommand{\cX}{\mathcal{X}}

\newcommand{\cZ}{\mathcal{Z}}

\newcommand{\rhoE}{\ensuremath{\idx{\rho}{E}}}

\newcommand{\chh}[5]{\ensuremath{H_{#1}^{#2}(\textnormal{#3}|\textnormal{#4})_{#5}}}

\newcommand{\chmineps}[3]{\chh{\textnormal{min}}{\eps}{#1}{#2}{#3}}
\newcommand{\chmineeps}[4]{\chh{\textnormal{min}}{#1}{#2}{#3}{#4}}

\newcommand{\chmax}[3]{\chh{\textnormal{max}}{}{#1}{#2}{#3}}
\newcommand{\chmaxeps}[3]{\chh{\textnormal{max}}{\eps}{#1}{#2}{#3}}
\newcommand{\chmaxeeps}[4]{\chh{\textnormal{max}}{#1}{#2}{#3}{#4}}

\theoremstyle{plain}
\newtheorem{lemma}{Lemma}
\newtheorem{theorem}[lemma]{Theorem}

\theoremstyle{definition}

\newcommand{\bfZkey}{\textbf{Z}}
\newcommand{\bfZkeyp}{\ensuremath{\textbf{Z}'}}
\newcommand{\bfXkey}{\textbf{X}}
\newcommand{\bfXest}{\hat{\textbf{X}}}
\newcommand{\bfXkeyp}{\ensuremath{\textbf{X}'}}
\newcommand{\Errpe}{\Lambda}
\newcommand{\errpe}{\lambda}

\newcommand{\leakEC}{\textnormal{leak}_{\textnormal{EC}}}
\newcommand{\ekr}{r}
\newcommand{\tn}[1]{\textnormal{#1}}
\newcommand{\Qtol}{Q_{\textnormal{tol}}}

\newcommand{\fkl}{r_{\textnormal{rel}}}
\newcommand{\ppass}{p_{\textnormal{pass}}}
\newcommand{\pabort}{p_{\textnormal{abort}}}
\newcommand{\e}{\epsilon}
\newcommand{\ec}{\epsilon_{\textnormal{cor}}}
\newcommand{\es}{\epsilon_{\textnormal{sec}}}
\newcommand{\er}{\epsilon_{\textnormal{rob}}}
\newcommand{\prot}{\boldsymbol{\Phi}}

\begin{document}

\title{Tight Finite-Key Analysis for Quantum Cryptography}
%

\author{Marco Tomamichel}
\email{marcotom.ch@gmail.com}
\affiliation{Institute for Theoretical Physics, ETH Zurich, 8093 Zurich,
Switzerland}

\author{Charles Ci Wen Lim}
\email{ciwen.lim@unige.ch}
\author{Nicolas Gisin}
\affiliation{Group of Applied Physics, University of Geneva, 1211 Geneva,
Switzerland}

\author{Renato Renner}
\affiliation{Institute for Theoretical Physics, ETH Zurich, 8093 Zurich,
Switzerland}

\begin{abstract}
Despite enormous theoretical and experimental progress in quantum cryptography,
the security of most current implementations of quantum key distribution is still not rigorously established. One significant problem is that the security of the final key strongly depends on the number, $M$, of signals exchanged between the legitimate parties|yet existing security proofs are often only valid asymptotically, for unrealistically large values of $M$. Another challenge is that most security proofs are very sensitive to small differences between the physical devices used by the protocol and the theoretical model used to describe them. Here, we show that these gaps between theory and experiment can be simultaneously overcome by using a recently developed proof technique based on the uncertainty relation for smooth entropies.
\end{abstract}

\maketitle

Quantum Key Distribution (QKD), invented by Bennett and
Brassard~\cite{bb84} and by Ekert~\cite{Ekert91}, can be considered
the first application of quantum information science, and commercial
products have already become
available. 
Accordingly, QKD has been an object of intensive study over
the past few years.  On the theory side, the security of several
variants of QKD protocols against general attacks has been
proved~\cite{lo99,sp00,biham06,mayers01,renner05,rennergisin05}. At
the same time, experimental techniques have reached a state of
development that enables key distribution at MHz rates over distances
of 100 km~\cite{takesue07,shields08,stucki09}.

Despite these developments, there is still a large gap between theory
and practice, in the sense that the security claims are based on
assumptions that are not (or cannot be) met by experimental
implementations. For example, the proofs often rely on theoretical
models of the devices (such as photon sources and detectors) that do
not take into account experimentally unavoidable imperfections 
(see~\cite{scarani09} for a discussion).
In this work, we consider \emph{prepare-and-measure} quantum key distribution protocols, like the original Bennett-Brassard 1984 (BB84) protocol~\cite{bb84}. Here, one party prepares quantum systems (e.g.\ the polarization degrees of freedom of photons) and sends them through an insecure quantum channel to another party who then measures the systems. In order to analyze the security of such protocols, the physical devices used by both parties to prepare and measure quantum systems are replaced by theoretical device models.
The goal, from a theory perspective, is to make these theoretical models as general as possible so 
that they can accommodate imperfect physical devices independently of their actual implementation. 
(This approach, in the context of \emph{entanglement-based} protocols, also led to the development
of device-independent quantum cryptography|see~\cite{haenggithesis10,masanes10} for recent results.)

Another weakness of many security proofs is the \emph{asymptotic resource assumption},
i.e., the assumption that an arbitrarily large number $M$ of signals
can be exchanged between the legitimate parties and used for the
computation of the final key.  This assumption is quite common in the
literature, and security proofs are usually only valid asymptotically
as $M$ tends to infinity. However, the asymptotic resource assumption
cannot be met by practical realizations|in fact, the key is often
computed from a relatively small number of signals ($M \ll
10^6$). This problem has recently received increased attention and
explicit bounds on the number of signals required to guarantee
security have been
derived~\cite{hayashi06-2,meyer06,inamori07,scarani08,bouman09,bratzik10,sheridan10}.

In this work, we apply a novel proof technique~\cite{tomamichel11} that allows us
to overcome the above difficulties. In particular, we derive almost tight bounds 
on the minimum value $M$ required to
achieve a given level of security. The technique is based on an
entropic formulation of the uncertainty relation~\cite{berta10} or,
more precisely, its generalization to smooth
entropies~\cite{tomamichel11}.  Compared to preexisting methods, our
technique is rather direct. It therefore avoids various estimates, including
the de Finetti Theorem~\cite{renner07} and the Post-Selection technique~\cite{christandl09}, that
have previously led to too pessimistic bounds.  Roughly speaking, our
result is a lower bound on the achievable key rate which deviates from
the asymptotic result (where $M$ is infinitely large) only by terms
that are caused by (probably unavoidable) statistical fluctuations in
the parameter estimation step. 
Moreover, we believe that the theoretical device model used for our 
security analysis is as general as possible 
for protocols of the prepare-and-measure type.

\section*{Results}


\subsection*{Security Definitions}

We follow the discussion of composable security~\cite{canetti01} and first take an abstract view on QKD protocols.
A QKD protocol describes the interaction between two players, Alice
and Bob. Both players can generate fresh randomness and 
have access to an insecure quantum channel as
well as an authenticated (but otherwise insecure) classical
channel. (Note that, using an authentication protocol, any insecure
  channel can be turned into an authentic channel. The authentication
  protocol will however use some key material, as discussed
  in~\cite{muller09}.)

The QKD protocol outputs a key, $\textbf{S}$, on Alice's side and an
estimate of that key, $\hat{\textbf{S}}$, on Bob's side. This key is
usually an $\ell$-bit string, where $\ell$, depends on the noise level
of the channel, as well as the security and correctness requirements
on the protocol. The protocol may also abort, in which case we set
$\textbf{S} = \hat{\textbf{S}} = \perp$.

In the following, we define what it means for a QKD protocol to be
\emph{secure}. Roughly speaking, the protocol has to (approximately)
satisfy two criteria, called \emph{correctness} and
\emph{secrecy}. These criteria are conditions on the probability
distribution of the protocol output, ${\textbf{S}}$ and
$\hat{\textbf{S}}$, as well as the information leaked to an adversary,
E. These depend, in general, on the attack strategy of the adversary,
who is assumed to have full control over the quantum channel
connecting Alice and Bob, and has access to all messages sent over the
authenticated classical channel.

  A QKD protocol is called \emph{correct} if, for any strategy of the
  adversary, $\hat{\textbf{S}} = \textbf{S}$.  It is called
  $\ec$-\emph{correct} if it is $\ec$-in\-dis\-tin\-guish\-able from a
  correct protocol.  In particular, a protocol is $\ec$-correct if
  $\Pr[\hat{\textbf{S}} \neq \textbf{S}] \leq \ec$.

In order to define the secrecy of a key, we consider the quantum state
$\idx{\rho}{\textbf{S}E}$ that describes the correlation between
Alice's classical key $\textbf{S}$ and the eavesdropper, E (for any
given attack strategy). A key is called $\Delta$-\emph{secret} from E
if it is $\Delta$-close to a uniformly distributed key that is
uncorrelated with the eavesdropper, i.e.\ if
\begin{align}
  \frac{1}{2} \normbig{\idx{\rho}{\textbf{S}E} - \idx{\omega}{\textbf{S}}
    \kron \rhoE}{1} \leq \Delta \,,
\end{align}
where $\idx{\omega}{\textbf{S}}$ denotes the fully mixed state on
$\textbf{S}$ and $\rhoE$ is the marginal state on the system E. For a motivation and discussion of this particular
secrecy criterion (in particular the choice of the norm) we refer
to~\cite{koenigrenner07}.
    
  A QKD protocol is called \emph{secret} if, for any attack strategy,
  $\Delta = 0$ whenever the protocol outputs a key. It is called
  $\es$-\emph{secret} if it is $\es$-indistinguishable from a secret
  protocol. In particular, a protocol is $\es$-secret if it outputs
  $\Delta$-secure keys with $(1-\pabort) \Delta \leq \es$, where
  $\pabort$ is the probability that the protocol aborts. (To
    see that this suffices to ensure $\es$-indistinguishability, note
    that the secrecy condition is trivially fulfilled if the protocol
    aborts.)

In some applications it is reasonable to consider correctness and
secrecy of protocols separately, since there may be different
requirements on the correctness of the key (i.e., that Bob's key
agrees with Alice's, implying that messages encrypted by Alice are
correctly decrypted by Bob) and secrecy. In fact, in many realistic
applications, an incorrect decoding of the transmitted data would be
detected so that the data can be resent. For such applications, $\ec$
may be chosen larger than $\es$.

However, secrecy of the protocol alone as defined above does not
ensure that Bob's key is secret from the eavesdropper as well. One is
thus often only interested in the overall security of the protocol
(which automatically implies secrecy of Bob's key).

  A QKD protocol is called \emph{secure} if it is correct and secret. It is called
  $\e$-\emph{secure} if it is $\e$-indistinguishable from a secure protocol. In particular, a protocol is $\e$-secure if it is $\ec$-correct and $\es$-secret with $\ec + \es \leq \e$.

Finally, the robustness, $\er$, is the probability that the protocol
aborts even though the eavesdropper is inactive. (More
  precisely, one assumes a certain channel model which corresponds to
  the characteristics of the channel in the absence of an
  adversary. For protocols based on qubits, the standard channel model
  used in the literature is the depolarizing channel. We also chose
  this channel model for our analysis in the discussion section, thus
  enabling a comparison to the existing results.) Note that a trivial
protocol that always aborts is secure according to the above
definitions, and a robustness requirement is therefore necessary. In
this work, we include the robustness $\er$ in our estimate for the
expected key rate (when the eavesdropper is inactive) and then
optimize over the protocol parameters to maximize this rate. 

\subsection*{Device Model}
\label{sec:model}

Recall that Alice and Bob are connected by an insecure quantum
channel. 
On one side of this channel, Alice controls a device allowing her to prepare quantum states in two bases, $\mathbb{X}$ and $\mathbb{Z}$. 
In an optimal scenario, the prepared states are qubits and the two bases are diagonal, e.g.\ $\mathbb{X} = \{ \ket{0}, \ket{1} \}$ and $\mathbb{Z} = \{ \ket{+}, \ket{-} \}$ with $\ket{\pm} := \big( \ket{0} \pm \ket{1} \big)/\sqrt{2}$. More generally, we characterize the quality of a source by its \emph{preparation quality}, $q$. The preparation quality|as we will see in the following|is the only device parameter relevant for our security analysis. It achieves its maximum of $q = 1$ if the prepared states are qubits and the bases are diagonal, as in the example above. In the following, we discuss two possible deviations from a perfect source and how they can be characterized in terms of~$q$.

Firstly, if the prepared states are guaranteed to be qubits, we characterize the quality of 
Alice's device by the maximum fidelity it allows between states prepared in the $\mathbb{X}$ basis 
and states prepared in the $\mathbb{Z}$ basis. Namely, we have $q = -\log \max \abs{\braket{\psi_
{x}}{\psi_{z}}}^{2}$, where the maximization is over all states $\psi_{x}$ and $\psi_{z}$ prepared 
in the $\mathbb{X}$ and $\mathbb{Z}$ basis, respectively. (In this work, $\log$ denotes the binary 
logarithm.) The maximum $q = 1$ is achieved if the basis states are prepared in diagonal bases, 
as is the case in the BB84 protocol.

In typical optical schemes, qubits are realized by polarization states of single 
photons.  An ideal implementation therefore requires a single-photon source in 
Alice's laboratory. In order to take into account sources that emit weak coherent light pulses instead, the analysis presented in this paper can be extended using photon tagging~\cite{lutkenhaus00} and decoy states~\cite{lo05}. This approach|although beyond the scope of the present article|can be incorporated into our finite-key analysis. 
(See also~\cite{hasegawa07,cai09,song11} for recent results on the finite-key analysis of such protocols.)

Secondly, consider a source that prepares states in the following way: The source produces two entangled particles and then sends out one of them while the other is measured in one of two bases. The choice of basis for the measurement decides whether the states are prepared in the $\mathbb{X}$ or $\mathbb{Z}$ basis. Together with the measurement outcome, which is required to be uniformly random for use in our protocol, this determines which of the four states is prepared. For such a source, the preparation quality is given by $q = -\log \max \norm{\sqrt{M_x} \sqrt{N_z}}{\infty}^2$, where $\{ M_x \}$ and $\{ N_z \}_z$ are the elements of the positive operator valued measurements (POVMs) that are used to prepare the state in the $\mathbb{X}$ and the $\mathbb{Z}$ basis, respectively.
If the produced state is that of two fully entangled qubits and the measurements are projective measurements in diagonal bases, we recover BB84 and $q = 1$~\cite{bennett92}. Sources of this type have recently received increased attention since they can be used as heralded single photon sources~\cite{pittman05,xiang10} and have applications in (device independent) quantum cryptography~\cite{gisin10,curty11,pitkanen11}.


On the other side of the channel, Bob controls a device allowing him
to measure quantum systems in two bases corresponding to $\mathbb{X}$
and $\mathbb{Z}$. We will derive security bounds that are valid
independently of the actual implementation of this device as long as
the following condition is satisfied: we require that the probability
that a signal is detected in Bob's device is independent of the basis
choices ($\mathbb{X}$ or $\mathbb{Z}$) by Alice and Bob. Note that
this assumption is necessary. In fact, if it is not satisfied (which
is the case for some implementations) a loophole arises that can be
used to eavesdrop on the key without being
detected~\cite{lydersen10}. (Remarkably, this assumption can
  be enforced device-independently: Bob simply substitutes a random
  bit whenever his device fails to detect Alice's signal. If this is
  done, however, the expected error rate may increase significantly.)

Finally, we assume that it is theoretically possible to devise an apparatus for Bob which delays all the measurements in the $\mathbb{X}$-basis until after parameter estimation, but produces the exact same measurement statistics as the actual device he uses. 
This assumption is satisfied if Bob's actual measurement device is memoryless. (To see this,
note that we could (in theorey) equip such a device with perfect quantum memory that stores
the received state until after the parameter estimation has been done.)
The assumption is already satisfied if the measurement statistics are unaffected when the memory of the actual device is reset after each measurement.
It is an open question whether this assumption can be further relaxed.

\subsection*{Protocol Definition}

We now define a family of protocols, $\prot[n, k, \ell, \Qtol, \ec,
\leakEC]$, which is parametrized by the \emph{block size}, $n$, the
number of bits used for parameter estimation, $k$, the \emph{secret
  key length}, $\ell$, the \emph{channel error tolerance}, $\Qtol$,
the required correctness, $\ec$, and the \emph{error correction
  leakage}, $\leakEC$. The protocol is asymmetric, so that the
number of bits measured in the two bases ($n$ bits in the $\mathbb{X}$
basis and $k$ bits in the $\mathbb{Z}$ basis) are not necessarily
equal~\cite{lo04}.

These protocols are described in Table~\ref{tb:protocol}.

\begin{table}
\begin{description}
	\item[State Preparation]
	The first four steps of the protocol are repeated for $i = 1, 2, \ldots, M$ 
	until the condition in the Sifting step is met.
	
	Alice chooses a basis $a_{i} \in \{ \mathbb{X},\, \mathbb{Z} \}$, where 
	$\mathbb{X}$ is chosen with probability $p_{x} = \big(1 + \sqrt{k/n}\big)^{-1}$ and 
	$\mathbb{Z}$ with probability $p_{z} = 1 - p_{x}$. (These probabilities are chosen in order to minimize the number $M$ of exchanged particles before Alice and Bob agree on the basis $\mathbb{X}$ for $n$ particles and on the basis $\mathbb{Z}$ for $k$ particles.)
  Next, Alice chooses a uniformly random bit $y_{i} \in \{0, 1\}$
	and prepares the qubit in a state of basis $a_{i}$, given by $y_{i}$. Alternatively, if the source is entanglement-based, Alice will ask it to prepare a state in the basis $a_{i}$ and record the output in $y_{i}$.
		
      \item[Distribution] Alice sends the qubit over the quantum
        channel to Bob. (Recall that Eve is allowed to arbitrarily
        interact with the system and we do not make any assumptions
        about what Bob receives.)
	
      \item[Measurement] Bob also chooses a basis, $b_{i} \in \{
        \mathbb{X},\, \mathbb{Z} \}$, with probabilities $p_{x}$ and
        $p_{z}$, respectively. He measures the system received from
        Alice in the chosen basis and stores the outcome in $y_{i}'
        \in \{0,1,\emptyset\}$, where `$\emptyset$' is the symbol
        produced when no signal is detected.
	
	\item[Sifting] Alice and Bob broadcast their basis choices over the classical
	channel. We define the sets $\mathcal{X} := \{ i : a_{i} = b_{i} = 
	\mathbb{X} \wedge y_i' \neq \emptyset\}$
	and $\mathcal{Z} := \{ i : a_{i} = b_{i} = \mathbb{Z} \wedge y_i' \neq \emptyset\}$. 
	The protocol repeats the first steps as long as either $\abs{\mathcal{X}} < n$ 
	or $\abs{\mathcal{Z}} < k$.
	
	\item[Parameter Estimation]
	Alice and Bob choose a random subset of size $n$ of $\mathcal{X}$
	and store the respective bits, $y_{i}$ and $y_{i}'$, into 
	\emph{raw key} strings 
	$\bfXkey$ and $\bfXkeyp$, respectively. 
	
	Next, they compute the average error 
	$\errpe := \frac{1}{\abs{\cZ}} \sum y_{i} \oplus y_{i}'$, where the sum is 
	over all $i \in \cZ$.
	The protocol aborts if $\lambda > \Qtol$.

	\item[Error Correction]
	An information reconciliation scheme that broadcasts at most 
	$\leakEC$ bits of classical error correction data is applied.
	This allows Bob to compute an estimate, $\bfXest$, of $\bfXkey$.
	
	Then, Alice computes a bit string (a hash) of length 
	$\lceil \log (1/\ec) \rceil$ by applying a random universal$_2$ hash function~\cite{carter79} to $\bfXkey$. She sends the choice of function and the hash to Bob. If the hash of $\bfXest$ disagrees with the hash of $\bfXkey$, the protocol aborts. 
		
	\item[Privacy Amplification]
	Alice extracts $\ell$ bits of secret key $\textbf{S}$ from 
	$\bfXkey$ using a random universal$_2$ hash function~\cite{bennett95,rennerkoenig05}. (Instead of
    choosing a universal$_2$ hash function, which requires at least $n$ bits of
    random seed, one could instead employ almost two-universal$_2$ hash
    functions~\cite{tomamichel10} or constructions based on Trevisan's
    extractor~\cite{portmann09}. These techniques allow for a reduction in the
    random seed length while the security claims remain almost unchanged.)
	The choice of function is communicated to Bob, who uses it to calculate
  $\hat{\textbf{S}}$.
\end{description}

\caption{Protocol Definition.}
\label{tb:protocol}
\end{table}

\subsection*{Security Analysis}

The following two theorems constitute the main technical result of our paper, stating that the protocols described above are both $\ec$-correct and $\es$-secure if the secret key length is chosen appropriately. Correctness is guaranteed  by the error correction step of the protocol, where a hash of Alice's raw key is compared with the hash of its estimate on Bob's side.
The following holds:
\begin{center}
\emph{The protocol $\prot[n, k, \ell, \Qtol, \ec, \leakEC]$ is $\ec$-correct.}
\end{center}

The protocols are $\es$-secure if the length of the extracted secret key does not exceed a certain length. Asymptotically for large block sizes $n$, the reductions of the key length due to finite statistics and security parameters can be neglected, and a secret key of length $\ell_{\max} = n (q - h(\Qtol)) - \leakEC$ can be extracted securely. Here, $h$ denotes the binary entropy function. Since our statistical sample is finite, we have to add to the tolerated channel noise 
a term $\mu \approx \sqrt{1/k \cdot \ln (1/\es)}$ that accounts for statistical fluctuations. Furthermore, the security parameters lead to a small reduction of the key rate logarithmic in $\ec$ and $\es$. The following theorem holds:

\noindent \emph{The protocol $\prot[n, k, \ell, \Qtol, \ec, \leakEC]$ using a source 
  with preparation quality 
  $q$ is $\es$-secret if the secret key length $\ell$ satisfies
  \begin{align}
    \label{eqn:keylength}
    \ell \leq 
    n \big(q - h (\Qtol + \mu) \big) - \leakEC - \log \frac{2}{\es^2 \ec} 
    \quad \textnormal{where} \quad
    \mu := \sqrt{\frac{n + k}{n k} \, \frac{k+1}{k} \, \ln 
    \frac{4}{\es}} \,.
  \end{align}
}

A sketch of the proof of these two statements follows in the methods section and a rigorous
proof of slightly more general versions of the theorems presented above can be found in Supplementary Material~$1$.

\section*{Discussion\label{sec:sim}}

In this section, we discuss the asymptotic behavior of our security bounds and
compare numerical bounds on the key rate for a finite number of exchanged
signals with previous results.
For this purpose, we assume that the quantum channel in the
absence of an eavesdropper can be described as a depolarizing channel with
\emph{quantum bit error rate} $Q$. (Note that this assumption is not needed for the security analysis of the previous section.)
The numerical results are computed for a perfect single-photon source, i.e.\ $q = 1$. Furthermore, finite detection efficiencies and channel losses are not factored into the key rates, i.e.\ the expected secret key rate calculated here can be understood as the expected key length per detected signal.

The efficiency of a protocol $\prot$ is characterized in terms of 
its \emph{expected secret key rate},
\begin{align}
	\label{eqn:ekr}
	\ekr(\prot, Q) := \big(1 - \er \big) \frac{\ell}{M(n, k)} \,,
\end{align}
where $M(n, k)$ is the expected number of qubits that need to be exchanged
until $n$ raw key bits and $k$ bits for parameter estimation are
gathered (see protocol description).

\begin{figure}[h!]
  \centering
  \includegraphics[width=13cm]{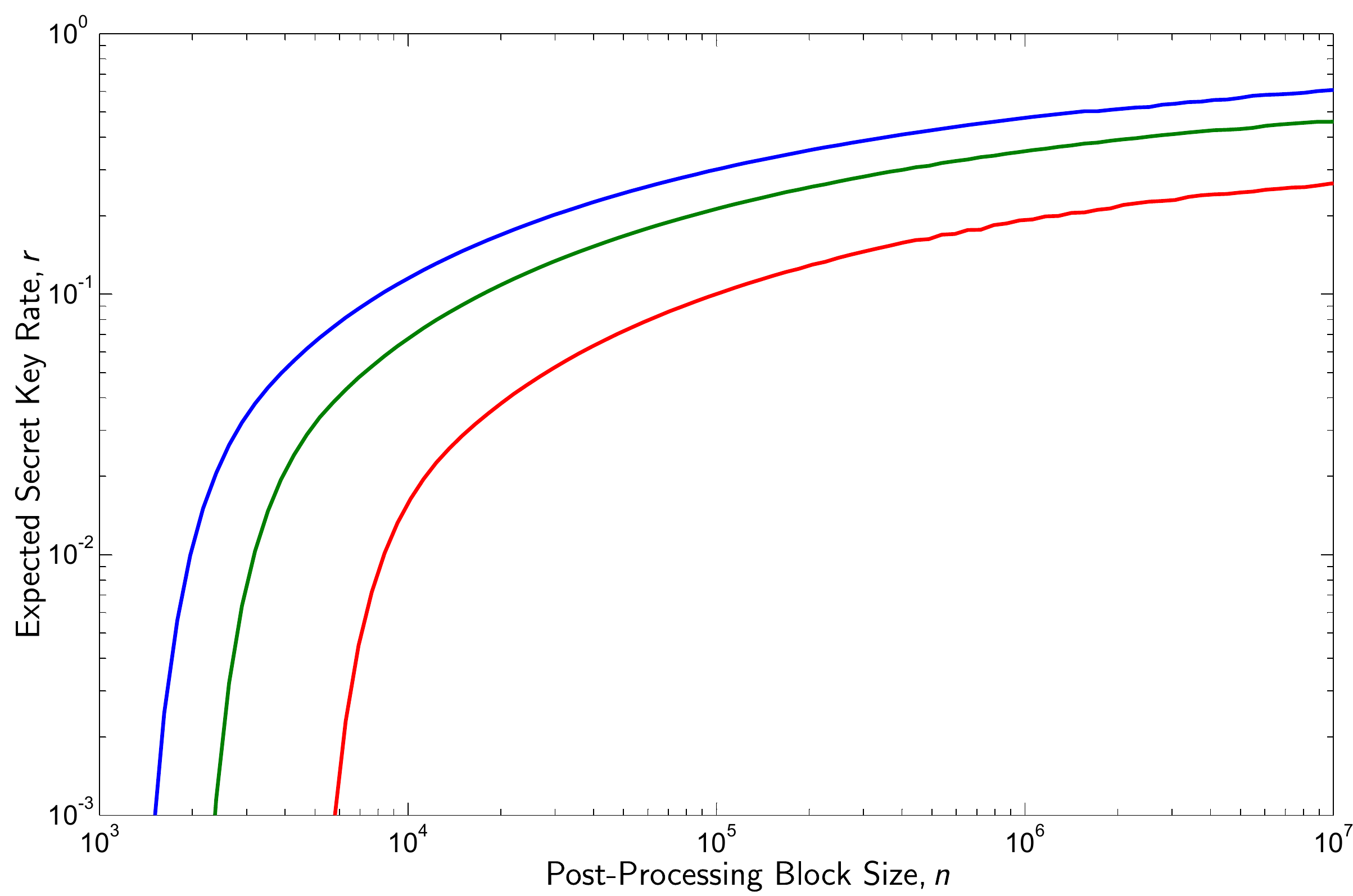}
  \caption{{\bf Expected Key Rate as Function of the Block Size.} 
  Plot of expected key rate $\ekr$ as a function of the block
    size $n$ for channel bit error rates $Q \in \{1\%, 2.5\%, 5\%\}$
    (from left to right). The security rate is fixed to $\epsilon/\ell
    = 10^{-14}$.}
  \label{fig:ours}
\end{figure}
 
Before presenting numerical results for the optimal expected key rates
for finite $n$, let us quickly discuss its asymptotic behavior for
arbitrarily large $n$. It is easy to verify that the key rate
asymptotically reaches $r_{\max}(Q) = 1 - 2 h(Q)$ for arbitrary
security bounds $\e > 0$.  To see this, note that error correction can
be achieved with a leakage rate of $h(Q)$ (see,
e.g.~\cite{cover91}). Furthermore, if we choose, for instance, 
$k$ proportional to
$\sqrt{n}$, the statistical deviation in~\eqref{eqn:keylength}, $\mu$,
vanishes and the ratio between the raw key length, $n$, and the
expected number of exchanged qubits, $M(n,k)$, approaches one as $n$
tends to infinity, i.e., $n/M(n,k) \to 1$.  This asymptotic rate is
optimal~\cite{rennerkraus05}.  Finally, the deviations of the key
length in~\eqref{eqn:keylength} from its asymptotic limit can be
explained as fluctuations that are due to the finiteness of the
statistical samples we consider and the error bounds we chose.
These terms are necessary for any finite-key analysis.
In particular, one expects a statistical deviation $\mu$ that scales with the inverse of the square root of the sample size $k$ as in~\eqref{eqn:keylength} from any statistical estimation of the error rate. In this sense our result is tight.

To obtain our results for finite block sizes $n$, we fix a security
bound $\e$ and define an optimized $\e$-secure protocol, $\prot^*[n,
\e]$, that results from a maximization of the expected secret key rate
over all $\e$-secure protocols with block size $n$.
For the purpose of this optimization, we assume an
error correction leakage of $\leakEC = \xi\,n\, h(\Qtol)$ with $\xi = 1.1$.
Moreover, we bound the robustness $\er$ by the probability that the measured security parameter exceeds $\Qtol$, which (for depolarizing channels) decays exponentially in $\Qtol - Q$. (Note that, for general quantum channels, the error rate in the $\mathbb{X}$ and $\mathbb{Z}$ bases may be different. Hence, the error correction leakage is in general not a function of $\Qtol$ but of the expected error rate in the $\mathbb{X}$ basis. Similarly, $\er$ generally is the sum of the robustness of parameter estimation as above and the robustness of the error correction scheme. In this discussion, the analysis is simplified since we consider a depolarizing channel, and, thus, the expected error rate is the same in both bases.)

In Figure~\ref{fig:ours}, we present the expected key rates $\ekr =
\ekr(\prot, Q)$ of the optimal protocols $\prot^*[n, \e]$ as a
function of the block size $n$.  These rates are given for a fixed
value of the \emph{security rate} $\eps / \ell$, i.e., the amount by
which the security bound $\eps$ increases per generated key bit. (In
other words, $\eps / \ell$ can be seen as the probability of key
leakage per key bit.)  The plot shows that significant key rates can
be obtained already for $n = 10^4$.

\begin{table}
\begin{center}
  \setlength{\tabcolsep}{9.5pt}
  \renewcommand{\arraystretch}{1.2}
  \begin{tabular}{ |c| c| c c c c c|}
    \hline
    $n$  &    				$Q$ $(\%)$ 	 & $\ekr$ $(\%)$  &   $\fkl$ $(\%)$&			$p_z$ $(\%)$  & 	$\Qtol$ $(\%)$  	& 	$\eps_{\tn{rob}}$ $(\%)$\\	
    \hline
    \multirow{2}{*}{$10^4$}& 	$1.0 $ 		& $11.7$   			& $14.0$ &      	$38.2$   		& 	$2.48$  		& 	$2.3$ \\	
 				 & 	$2.5 $ 	 	& $6.8$   			& $10.4$&      		$43.0$   		& 	$3.78$  		& 	$3.0$ \\	
    \hline
    \multirow{2}{*}{$10^5$}& 	$1.0 $ 	 	& $30.4$   			& $36.4$&  		$22.0$   		& 	$2.14$  		& 	$0.8$ \\	
			 	& 	$2.5 $ 	 	& $21.5$   			& $32.6$&     		$23.3$   		& 	$3.58$  		& 	$1.0$ \\	
    \hline
    \multirow{2}{*}{$10^6$}& 	$1.0 $ 		& $47.8$   			& $57.1$ &      	$12.5$   		& 	$1.73$  		& 	$0.6$ \\	
 				& 	$2.5 $ 	 	& $35.7$  			& $53.9$ &      	$13.7$   		& 	$3.21$  		& 	$0.7$ \\	
    \hline
  \end{tabular}
  \caption{Optimized parameters for a given security rate
    $\epsilon/\ell = 10^{-14}$.  The column labeled $\fkl$ shows the
    deviation of the expected secret key rate from the corresponding
    asymptotic value, i.e., $ \fkl := \ekr/(1 - 2h(Q))$.}
  \label{tbl:num}
\end{center}
\end{table}

In Table~\ref{tbl:num}, we provide selected numerical results for the
optimal protocol parameters that correspond to block sizes $n =
\{10^{4}, 10^{5}, 10^{6}\}$ and quantum bit error rates $Q \in \{ 1\%,
2.5\% \}$. These block sizes exemplify current hardware limitations in
practical QKD systems.

\begin{figure}[h!]
  \centering
  \includegraphics[width=13cm]{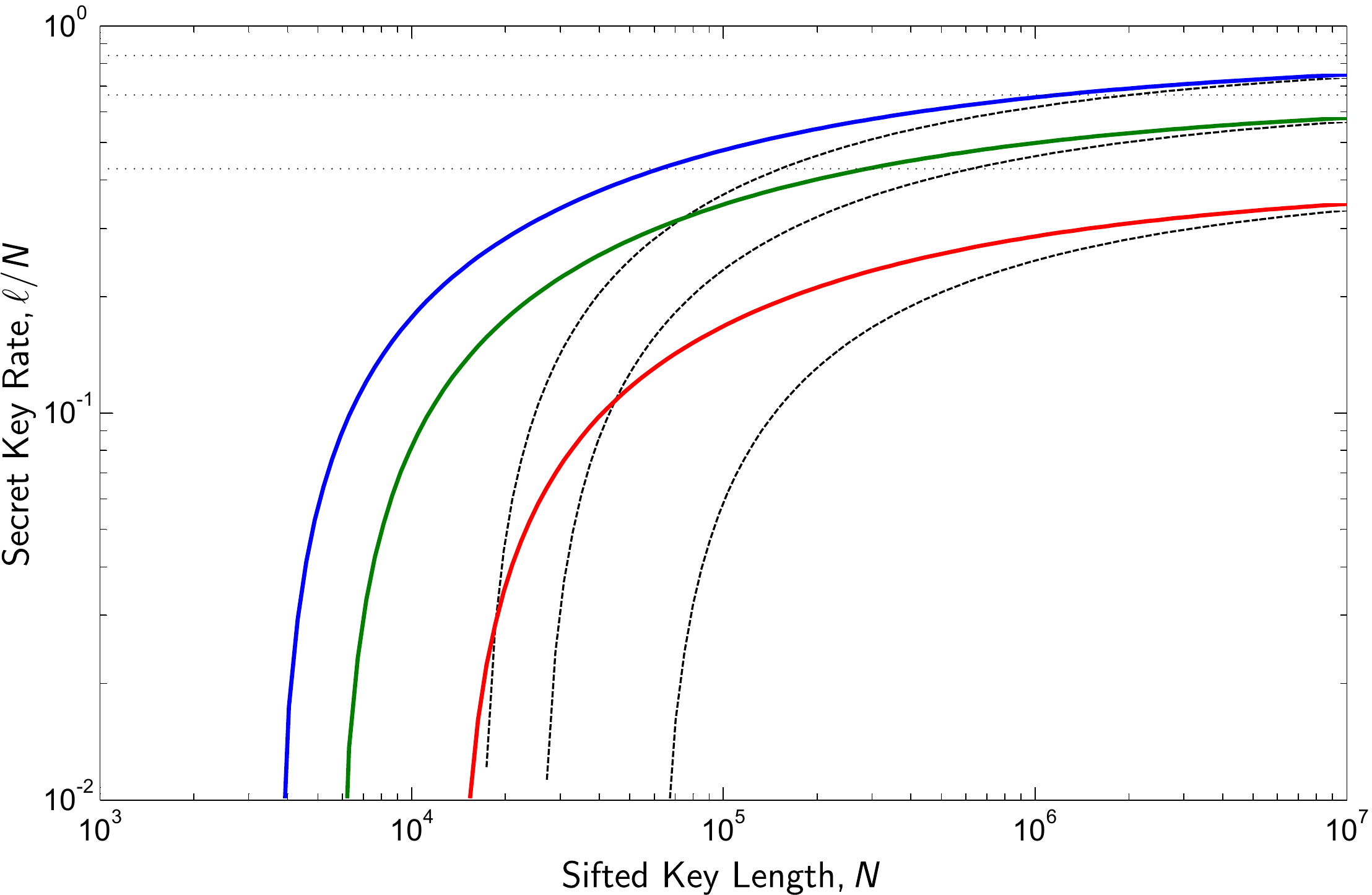}
  \caption{ {\bf Comparison of Key Rate with Earlier Results.} 
    The plots show the rate $\ell/N$ as a function of the
    sifted key size $N = n+k$ for various channel bit error rates $Q$
    (as in Fig.~\ref{fig:ours}) and a security bound of $\e =
    10^{-10}$.  The (curved) dashed lines show the rates that can be proven
    secure using~\cite{scarani08}. The horizontal dashed lines indicate the asymptotic rates for $Q \in \{1\%, 2.5\%, 5\%\}$ (from top to bottom).}
  \label{fig:comp}
\end{figure}

In Figure~\ref{fig:comp}, we compare our optimal key rates with the
maximal key rates that can be shown secure using the finite key
analysis of Scarani and Renner~\cite{scarani08}. For
  comparison with previous work, we plot the rate $\ell/n$, i.e.\ the
  ratio between key length and block size, instead of the expected
  secret key rate as defined by Eq.~\eqref{eqn:ekr}.  We show a major
improvement in the minimum block size required to produce a provably
secret key.  The improvements are mainly due to a more direct
evaluation of the smooth min-entropy via the entropic uncertainty
relation and the use of statistics optimized specifically to the
problem at hand (cf.\ Supplementary Note~$2$). 

In conclusion, this article gives tight finite-key bounds for secure quantum key distribution with an asymmetric 
BB84 protocol. Our novel proof technique, based on the uncertainty principle, offers a conceptual 
improvement over earlier proofs that relied on a tomography of the state shared between Alice and 
Bob. Most previous security proofs against general adversaries, e.g.~\cite{renner05,scarani08,sheridan10,bratzik10}, 
are arranged in two steps: An analysis of the security against adversaries restricted to collective attacks and a lifting of this argument to general attacks. The lifting is often possible without a 
significant loss in key rate 
using modern techniques~\cite{renner07,christandl09}; hence, the main difference lies in the 
first part. In security proofs against collective attacks Alice and Bob usually do tomography on 
their shared state, i.e., they characterize the density matrix of their shared state. 
Since the eavesdropper can be assumed to hold a purification of this state, it is then possible to bound the von Neumann entropy of the eavesdropper on Alice's measurement result. The min-entropy of the eavesdropper is in turn bounded using the quantum asymptotic equipartition property~\cite{renner05,tomamichel08}, introducing a penalty scaling with $1/\sqrt{n}$ on the key rate. 
(A notable exception is~\cite{bratzik10}, where the min-entropy is bounded directly from the results of tomography.)

In contrast, our approach bounds the min-entropy directly and does not require us to do tomography on the state shared between Alice and Bob. In fact, we are only interested in one correlation (between $\bfZkey$ and $\bfZkeyp$) and, thus, our statistics can be produced more efficiently. (Note, however, that this is also the reason why our approach does not reach the asymptotic key rate for the $6$-state protocol~\cite{bruss98}. There, full tomography puts limits on Eve's information that go beyond the uncertainty relation in~\cite{tomamichel11}.) 
Finally, since our considerations are rather general, we believe that they can be extended to other QKD protocols.

\section*{Methods\label{sec:proof}}

\subsection*{Correctness}

The required correctness is ensured in the error correction step of the protocol, when Alice and Bob compute and evaluate a random hash function of their keys. If these hash values disagree, the protocol aborts and both players output empty keys. (These keys are trivially correct.) Since arbitrary errors in the key will be detected with high probability when the hash values are compared~\cite{carter79}, we can guarantee that Alice's and Bob's secret keys are also the same with high probability.

\subsection*{Secrecy}

In order to establish the secrecy of the protocols, we 
consider a \emph{gedankenexperiment} in which Alice and Bob, after
choosing a basis according to probabilities $p_{x}$ and $p_{z}$ as
usual, prepare and measure everything in the $\mathbb{Z}$ basis. We
denote the bit strings of length $n$ that replace the raw keys
$\bfXkey$ and $\bfXkeyp$ in this hypothetical protocol as $\bfZkey$
and $\bfZkeyp$, respectively.
The secrecy then follows from the fact that, if Alice has a choice of encoding a string of $n$ uniform bits in either
the $\mathbb{X}$ or $\mathbb{Z}$ basis, the following holds: the better Bob is able to estimate Alice's string if she prepared in the $\mathbb{Z}$ basis, the worse Eve is able to guess Alice's string if she prepared in the $\mathbb{X}$ basis. This can be formally expressed in terms of an uncertainty relation for smooth entropies~\cite{tomamichel11},
\begin{align}
	\label{eqn:ucr}
	\chmineeps{\eps}{\bfXkey}{E}{} + \chmaxeeps{\eps}{\bfZkey}{\bfZkeyp}{} \geq n q \, ,
\end{align}
where $\eps \geq 0$ is called a smoothing parameter and $q$, as we will see below, is the
preparation quality defined previously. The smooth min-entropy, $\chmineps{\bfXkey}{E}{}$, introduced in~\cite{renner05}, characterizes the average probability that Eve guesses $\mathbf{X}$ correctly using her
optimal strategy with access to the correlations stored in her quantum
memory~\cite{koenig08}. The smooth max-entropy, $\chmaxeps{\bfZkey}{\bfZkeyp}{}$,
corresponds to the number of additional bits that are needed in order to reconstruct the value of $\bfZkey$ using $\bfZkeyp$ up to a failure probability $\eps$~\cite{renesrenner10}. For precise mathematical definitions of the smooth min- and max-entropy,
we refer to~\cite{tomamichel09}.

The sources we consider in this article are either a) qubit sources or b) sources that create {BB84}-states by measuring part of an entangled state. In case b), a comparison with~\cite{tomamichel11} reveals that the bound on the uncertainty is given by $-\log c$, where $c$ is the overlap of the two measurement employed in the source. For general POVMs, $\{ M_x \}$ for preparing in the $\mathbb{X}$ basis and $\{ N_z \}$ for preparing in the $\mathbb{Z}$ basis, this overlap is given by $c = \max \norm{\sqrt{M_x} \sqrt{N_z}}{\infty}^2$. This justifies the definition of the preparation quality $q = -\log c$ for such sources.
In case a), the preparation process can be purified into an entanglement-based one of the type above. To see this, simply consider a singlet state between two qubits and projective measurements on the first qubit. It is easy to verify that the overlap of the prepared states in the two bases is equal to the overlap of the two projective measurements used to prepare them. Hence, the preparation quality of this source is given by $q = -\log c$, where $c$ is the maximum overlap of the prepared states.

Note that|in the \emph{gedankenexperiment} picture|the observed average error, $\lambda$, is calculated from $k$ measurements sampled at random from $n + k$ measurements in the $\mathbb{Z}$ basis.
Hence, if $\lambda$ is
small, we deduce that, with high probability, $\bfZkey$ and
$\bfZkeyp$ are highly correlated and, thus, 
$\chmaxeps{\bfZkey}{\bfZkeyp}{}$ is small. In fact, since the protocol aborts if $\lambda$ exceeds $\Qtol$, the
following bound on the smooth max-entropy (conditioned on the correlation test passing) holds:
\begin{align}
  \label{eqn:pe}
	\chmaxeeps{\eps}{\bfZkey}{\bfZkeyp}{} 
	\leq n  h\big(\Qtol + \mu \big) ,
\end{align}
where $\mu$ takes into account statistical fluctuations and depends on the security parameter via $\eps$. Eq.~\eqref{eqn:pe} is shown in Supplementary Note $2$ using an upper bound by Serfling~\cite{serfling74} on the probability that the average error on the sample, $\lambda$, deviates by more than $\mu$ from the
average error on the total string. (See also~\cite{vanlint99}.)

In addition to the uncertainty relation, our analysis
employs the \emph{Quantum Leftover Hash Lemma}~\cite{renner05,tomamichel10}, which
gives a direct operational meaning to the smooth min-entropy. It
asserts that, using a random universal$_2$ hash function, it is
possible to extract a $\Delta$-secret key of length $\ell$ from
$\bfXkey$, where
\begin{align}
	\label{eqn:pa}
	\Delta = 2\eps + \frac{1}{2} \sqrt{2^{\ell - \chmineeps{\eps}{\bfXkey}{E$'$}{}}} \, .
\end{align}
Here, E$'$ summarizes all information Eve learned about $\bfXkey$
during the protocol|including the classical communication sent by
Alice and Bob over the authenticated channel. For the protocol discussed here, a maximum of $\leakEC + \big\lceil\log (1/\ec)\big\rceil$ bits of information about $\bfXkey$ are revealed to the eavesdropper during the protocol. Hence, using a chain rule for smooth min-entropies, we can relate the smooth min-entropy prior to the classical post-processing, $\chmineps{\bfXkey}{E}{}$, with the min-entropy before privacy amplification, $\chmineeps{\eps}{\bfXkey}{E$'$}{}$ as follows.
\begin{align}
 \chmineeps{\eps}{\bfXkey}{E$'$}{} \geq \chmineeps{\eps}{\bfXkey}{E}{} - \leakEC - \log \frac{2}{\ec} \, .
\end{align}
Collecting the bounds on the smooth entropies we got from the uncertainty relation,~\eqref{eqn:ucr}, and the parameter estimation,~\eqref{eqn:pe}, we further find that
\begin{align}
  \label{eqn:min-bound-last}
 \chmineeps{\eps}{\bfXkey}{E$'$}{} \geq n \big( q - h(\Qtol + \mu) \big) - \leakEC - \log \frac{2}{\ec} \,.
\end{align}

Combining this with the Quantum Leftover Hashing Lemma~\eqref{eqn:pa} and using the bound on the key length given in Eq.~\eqref{eqn:keylength}, we get
\begin{align}
  \label{eqn:safekey}
  \Delta \leq 2\eps + \frac{1}{2} \sqrt{2^{\ell - 
  \chmineeps{\eps}{\bfXkey}{E$'$}{}}} \leq 2\eps + \frac{\es}{2} \, .
\end{align}
Finally, the protocol is $\es$-secret if we choose $\eps$ proportional to $\es$ and 
sufficiently small.

\section*{Acknowledgments}

We would like to thank Silvestre Abruzzo, Normand Beaudry, Dagmar Bruss, Hermann
Kampermann, Markus Mertz, Norbert L\"utkenhaus, Christoph Pacher,
Momtchil Peev, Hugo Zbinden and Christopher Portmann for valuable comments and stimulating
discussions. We acknowledge support from the National Centre of
Competence in Research QSIT, the Swiss NanoTera project QCRYPT, the
Swiss National Science Foundation (grant no.~200021-119868), and the
European Research Council (grant No.~258932 and No.~227656).

\appendix

\renewcommand{\theequation}{S\arabic{equation}}
\setcounter{equation}{0}

\section*{Supplementary Note $1$: Finite Key Analysis}

The required correctness is ensured during the error correction step of any protocol in the family. There, Alice and Bob compute and compare a hash of length $\lceil \log (1/\ec) \rceil$ by applying a random universal$_2$ hash function to their keys, $\bfXkey$ and $\bfXest$. If the hash values disagree, the protocol aborts.

\begin{theorem}
  \label{thm:cor}
  The protocol $\prot[n, k, \ell, \Qtol, \ec, \leakEC]$ is $\ec$-correct. 
\end{theorem}

\begin{proof}
The defining property of such universal$_2$ families of hash functions~\cite{carter79}
is the fact that the probability with which $F(\bfXkey)$ and $F(\bfXest)$ coincide\,|\,if $\bfXkey$ and
$\bfXest$ are different and the hash function, $F$, is chosen uniformly at random from the family\,|\,is
at most $2^{-\lceil \log (1/\ec) \rceil}$.
Since the protocol aborts if the hash values calculated from $\bfXkey$ and
$\bfXest$ after error correction do not agree, it is thus ensured that $\Pr [
\textbf{S} \neq \hat{\textbf{S}} ] \leq \Pr [ \bfXkey \neq \bfXest ] \leq 2^{-\lceil \log (1/\ec) \rceil} \leq \ec$.
\end{proof}

In order to prove the security of the protocols, we
consider a \emph{gedankenexperiment} in which Alice and Bob, after
choosing a basis according to probabilities $p_{x}$ and $p_{z}$ as
before, prepare and measure everything in the $\mathbb{Z}$ basis. We
denote the bit strings of length $n$ that replace the raw keys
$\bfXkey$ and $\bfXkeyp$ in this hypothetical protocol as $\bfZkey$
and $\bfZkeyp$, respectively.

Security is now based on the observation that, if
Alice has a choice of encoding a string of $n$ uniform bits in either
the $\mathbb{X}$ or $\mathbb{Z}$ basis, then the following holds: the better Bob is able to estimate Alice's string if she prepared in the $\mathbb{Z}$ basis, the worse Eve is able to guess Alice's string if she prepared in the $\mathbb{X}$ basis.
This can be formally expressed in terms of an uncertainty
relation for smooth entropies~\cite{tomamichel11}
\begin{align}
	\label{eqn:ucr}
	\chmineeps{\eps'}{\bfXkey}{E}{} + \chmaxeeps{\eps'}{\bfZkey}{B}{} \geq n q \, ,
\end{align}
where $\eps' \geq 0$ is a smoothing parameter and $q$, as we see below, is the
preparation quality defined in the main text. 
The smooth min-entropy,
$\chmineps{\bfXkey}{E}{}$, introduced in~\cite{renner05},
characterizes the
average probability that Eve guesses $\bfXkey$ correctly using her
optimal strategy with access to the correlations stored in her quantum
memory~\cite{koenig08}.
The smooth max-entropy, $\chmaxeps{\bfZkey}{B}{}$,
is a measure of the correlations between $\bfZkey$ and Bob's data. For
precise mathematical definitions of the smooth min- and max-entropy,
we refer to~\cite{tomamichel09}.

The sources we consider in this article are either a) qubit sources or b) sources that create {BB84}-states by measuring part of an entangled state. In the latter case, a comparison with~\cite{tomamichel11}
reveals that the bound on the uncertainty is given by $-\log c$, where $c$ is the overlap of the two measurement employed in the source. For general POVMs, $\{ M_x \}$ for preparing in the $\mathbb{X}$ basis and $\{ N_z \}$ for preparing in the $\mathbb{Z}$ basis, this overlap is given by $c = \max \norm{\sqrt{M_x} \sqrt{N_z}}{\infty}^2$. This justifies the definition of the preparation quality $q = -\log c$ (as defined in Section~I.B of the main text) for such sources.

On the other hand, if the prepared state is guaranteed to be a qubit state, the preparation process can be purified into an entanglement-based one of the type above. To see this, simply consider a singlet state between two qubits and projective measurements on the first qubit. It is easy to verify that the overlap of the prepared states in the two bases is equal to the overlap of the two projective measurements used to prepare them. Hence, the preparation quality of this source is given by $q = -\log c$, where $c$ is the maximum overlap of the prepared states (as defined in Section~I.B of the main text).

Apart from the uncertainty relation~\eqref{eqn:ucr}, our analysis
employs the Quantum Leftover Hash Lemma~\cite{tomamichel10}
which
gives a direct operational meaning to the smooth min-entropy. It
asserts that, using a random universal$_2$ hash function, it is
possible to extract a $\Delta$-secret key of length $\ell$ from
$\bfXkey$, where
\begin{align}
	\label{eqn:pa}
	\Delta = \min_{\eps'} \frac{1}{2} \sqrt{2^{\ell - \chmineeps{\eps'}{\bfXkey}{E$'$}{}}} + 2\eps' \, .
\end{align}
Here E$'$ summarizes all information Eve learned about $\bfXkey$
during the protocol\,|\,including the classical communication sent by
Alice and Bob over the authenticated channel. Furthermore, the
extracted secret key is independent of the randomness that is used to
choose the hash function.

The following theorem gives a sufficient condition for which a
protocol $\prot$ using a source with preparation quality $q$ is
$\es$-secret.  The minimum value $\es$ for which it is $\es$-secret is
called the \emph{secrecy of the protocol} and is denoted by
$\es(\prot,q)$.

\begin{theorem}
  \label{thm:sec}
  The protocol $\prot[n, k, \ell, \Qtol, \ec, \leakEC]$ using a source 
  with preparation quality 
  $q$ is $\es$-secret for some $\es > 0$ if $\ell$ satisfies\footnote{Here, $h$ is a truncated binary entropy function, i.e.\ $h: x \mapsto - x \log x - (1-x) \log (1-x)$ if $x \leq 1/2$ and $1$ otherwise.}
  \begin{align}
    \label{eqn:keylength}
    \ell \leq \max_{\eps, \bar{\eps}} 
    \bigg\lfloor 
    n \Big(q - h \big(\Qtol + \mu(\eps)\big)\Big) - 
    2 \log \frac{1}{2\,\bar{\eps}} - \leakEC - \log \frac{2}{\ec} 
    \bigg\rfloor 
    ,
  \end{align}
  where we optimize over $\eps > 0$ and $\bar{\eps} > 0$ s.t.\ 
  $2\eps + \bar{\eps} \leq \es$ and
  \begin{align}
    \mu(\eps) := \sqrt{\frac{n + k}{n k} \, \frac{k+1}{k} \, \ln 
    \frac{1}{\eps}} \, .
  \end{align}
\end{theorem}

\begin{proof}
  In the \emph{gedankenexperiment} picture described above, $\Lambda$ is
  a random variable calculated from at least $k$ measurements sampled at random from $n
  + k$ measurements in the $\mathbb{Z}$ basis. Hence, if $\Lambda$ is
  small, we deduce that, with high probability, $\bfZkey$ and
  $\bfZkeyp$ are highly correlated and $\chmax{\bfZkey}{\bfZkeyp}{}$
  is small. This is elaborated in
  Lemma~\ref{lemma:stat}, where it is shown that, conditioned on the
  event that the correlation test passed ($\Lambda \leq \Qtol$), the
  following bound on the smooth max-entropy holds,
\begin{align}
	\chmaxeeps{\eps'}{\bfZkey}{\bfZkeyp}{\rho} 
	\leq n  h\big(\Qtol + \mu(\eps)\big) \,,
\end{align}
where $\eps' = \eps/\sqrt{\ppass}$ and $\ppass \geq 1 - \pabort$ is the
probability that the correlation test passes. Here, $\rho$ is the state of the system  conditioned on the event that the correlation test passed. More precisely, we consider the state $\rho_{\mathbf{AB}\textnormal{E}}$ of the $n$ systems shared between Alice and Bob as well as Eve's information. Moreover, the classical joint probability distributions $\rho_{\bfXkey\bfXkeyp}$ and $\rho_{\bfZkey\bfZkeyp}$ are induced by the respective measurement on $\mathbf{A}$ and $\mathbf{B}$.
(Note that these states are well-defined since, by assumption, we know that the measurement of the $n$ bits used for key generation can be postponed until after parameter estimation.)

We now apply the uncertainty relation, $\chmineeps{\eps'}{\bfXkey}{E}{\rho} \geq n  q
- \chmaxeeps{\eps'}{\bfZkey}{\bfZkeyp}{\rho}$, on this state to find a lower bound on the
min-entropy that Eve has about Alice's bits prepared in the $\mathbb{X}$ basis.
Since a maximum of $\leakEC + \big\lceil\log (1/\ec)\big\rceil \leq
\leakEC + \log (2/\ec)$ bits of information about $\bfXkey$ are revealed
during error correction, we find\footnote{Formally, this requires use of the
chain rule $\chmineps{\bfXkey}{EC}{} \geq \chmineps{\bfXkey}{E}{} - \log
\abs{\textnormal{C}}$, where C is any classical information about $\bfXkey$.}
\begin{align}
  \chmineeps{\eps'}{\bfXkey}{E$'$}{\rho} &\geq \chmineeps{\eps'}{\bfXkey}{E}{\rho} - \leakEC - 
    \log \frac{2}{\ec} \\
  &\geq n q - \chmaxeeps{\eps'}{\bfZkey}{\bfZkeyp}{\rho} - \leakEC - \log \frac{2}{\ec} \\
  &\geq n \Big( q - h\big(\Qtol + \mu(\eps)\big) \Big) - \leakEC - \log \frac{2}{\ec}\,.
\end{align}
Thus, combining this with~\eqref{eqn:pa} and using the proposed key length~\eqref{eqn:keylength}, we find, for all $\eps$ and $\bar{\eps}$,
\begin{align}
  \Delta \leq 2\eps' + \frac{1}{2} \sqrt{2^{\ell - 
  \chmineeps{\eps'}{\bfXkey}{E$'$}{\rho}}} \leq 2\eps' + \bar{\eps} \, .
  \label{eqn:safekey}
\end{align}
The security of the protocol now follows since $(1 - \pabort) \Delta \leq 2\eps + \bar{\eps} \leq \es$.
\end{proof}

\section*{Supplementary Note $2$: Statistics}

This section covers the statistical analysis of the classical data collected during the run of the BB84-type protocols described in this work. A more general framework for such an analysis can be found in \cite{bouman09}

We use the notation of the previous sections and define $N := n + k$. The fraction of bits 
that are used for parameter estimation is denoted as $\nu$, i.e.\ $k = \nu N$ and $n = (1 - \nu) N$.

\newcommand{\Errtot}{\Lambda_{\textnormal{tot}}}
\newcommand{\Errkey}{\Lambda_{\textnormal{key}}}
\newcommand{\Errpefull}{\Lambda_{\textnormal{pe}}}
\newcommand{\errtot}{\lambda_{\textnormal{tot}}}
\newcommand{\errkey}{\lambda_{\textnormal{key}}}
\newcommand{\bfZtot}{\ensuremath{\textbf{Z}_{\textnormal{tot}}}}
\newcommand{\bfZtotp}{\ensuremath{\textbf{Z}'_{\textnormal{tot}}}}
\newcommand{\bfZpe}{\ensuremath{\textbf{Z}_{\textnormal{pe}}}}
\newcommand{\bfZpep}{\ensuremath{\textbf{Z}'_{\textnormal{pe}}}}
\newcommand{\bfZkeyfull}{\ensuremath{\textbf{Z}_{\textnormal{key}}}}
\newcommand{\bfZkeypfull}{\ensuremath{\textbf{Z}'_{\textnormal{key}}}}

The statistical analysis is based on a \emph{gedankenexperiment},
where Alice and Bob measure all $N$ states with $i \in \cX \cup \cZ$
in the control basis, $\mathbb{Z}$, resulting in strings $\bfZtot$ and
$\bfZtotp$ for Alice and Bob, respectively. The following random
variables are of interest to us. The relative Hamming distance between
Alice's and Bob's bit-string is defined as $\Errtot = \frac{1}{N}
\Abs{ \bfZtot \oplus \bfZtotp }$, where $\abs{\cdot}$ denotes the
Hamming weight. Similarly, $\Errpe = \Errpefull$ denotes the relative
Hamming distances between the random subsets $\bfZpe$ of $\bfZtot$ and
$\bfZpep$ of $\bfZtotp$ used for parameter estimation. Finally,
$\Errkey$ is the relative Hamming distance between the remainders of
the strings, denoted $\bfZkey = \bfZkeyfull$ and $\bfZkeyp =
\bfZkeypfull$. Clearly,
$$ \Errtot = \nu \Errpe + (1 - \nu) \Errkey \, . $$


The $k$ bits used for parameter estimation are chosen at random from $N$ bits. Hence, 
if we fix $\Errtot = \errtot$ for the moment, the random variables $\Errpe$ and $\Errkey$ 
can be seen as emanating from sampling without replacement. We apply the bound~[54]
%
\begin{align}\label{eqn:serfling}
  \Pr \big[ \Errkey \geq \errtot + \delta \,|\, \Errtot = \errtot \big] 
  \leq e^{- 2 \frac{n N}{k+1} \delta^2} \, .
\end{align}
We now derive a bound on the probability that $\Errkey$ exceeds $\Errpe$ by more than a constant $\mu$ conditioned on the event that we passed the correlation test. (Note that, while $\Errpe$ is accessible during the protocol, $\Errkey$ is the quantity we are actually interested in.) We find, using Bayes' theorem,
\begin{align}
  &\Pr \big[ \Errkey \geq \Errpe + \mu\,|\textnormal{``pass''}\big] 
  \leq \frac{1}{\ppass} \Pr \big[ \Errkey \geq \Errpe + \mu \big] \, ,
\end{align}
where we keep $\ppass = \Pr[\textnormal{``pass''}] = \Pr[\Lambda \leq \Qtol]$ as a parameter and further bound
\begin{align}
  &\Pr \big[ \Errkey \geq \Errpe + \mu \big] = \Pr \big[ \Errkey
    \geq \Errtot +\nu \mu \big] \\
  &\quad = \sum_{\errtot} \Pr \big[ \Errtot = \errtot \big]\, \Pr \big[ 
 	  \Errkey \geq \errtot + \nu \mu \,|\, \Errtot = \errtot \big] 
	  \leq e^{- 2 \frac{k n}{N} \frac{k}{k + 1} \mu^2 } \, .
\end{align}
We used~\eqref{eqn:serfling} to bound each summand individually.
Finally, defining $\eps := e^{-\frac{kn}{N} \frac{k}{k+1} \mu^2}$, we write
\begin{align}
  \label{eqn:bbkey}
  \Pr \big[\Errkey \geq \Errpe + \mu\,|\textnormal{``pass''}\big] 
  \leq \frac{\eps^2}{\ppass} \,.
\end{align}


The above result can be used to bound the uncertainty Bob has about Alice's measurement outcomes in the $\mathbb{Z}$-basis, as expressed using the smooth max-entropy of $\bfZkey$ given $\bfZkeyp$ and $\Errpe$. The entropy is evaluated for the probability distribution conditioned on the event that the correlation test passed, which we denote $\mathds{P}_{\bfZkey\bfZkeyp\Errpe}(\textbf{z}, \textbf{z}', \errpe) = \Pr[ \bfZkey = \textbf{z} \wedge \bfZkeyp = \textbf{z}' \wedge \Errpe = \errpe\,|\textnormal{``pass''} ]$.

\begin{lemma}
  \label{lemma:stat}
  Let $\eps > 0$. Then
  \begin{align}
    \chmaxeeps{\eps'}{\bfZkey}{\bfZkeyp}{} 
    \leq n h 
      \big( \Qtol + \mu \big) \, ,
    \quad \textrm{where} \quad \eps' := \frac{\eps}{\sqrt{\ppass}} 
    \quad \textrm{and} \quad \mu := \sqrt{\frac{N}{n k} \, \frac{k+1}{k} \ln 
	  \frac{1}{\eps} } \, .
	\end{align}
\end{lemma}

\begin{proof}
  According to~\eqref{eqn:bbkey}, the probability that $\Errkey$ 
  exceeds $\Errpe$ by more than $\mu$ is bounded. In fact, we can find
  a probability distribution,
\begin{align}
  \mathds{Q}_{\bfZkey\bfZkeyp\Errpe}({\bf z}, {\bf z}', \errpe) := 
    \left\{  \begin{array}{ll} \frac{\mathds{P}_{\bfZkey\bfZkeyp\Errpe}({\bf z},{\bf z}',
    \errpe)}{\Pr[\Errkey < \Errpe + \mu\,|\textnormal{``pass''}]}
	& \ \textnormal{if}\ \
    \errkey({\bf z}, {\bf z}') < \errpe + \mu \\ 
    0 & \ \textnormal{else} \end{array} \right. \,,
\end{align}
  which is $\eps'$-close to $\mathds{P}_{\bfZkey\bfZkeyp\Errpe}$ in terms of the purified distance. To 
  see this, note that the fidelity between the two distributions satisfies 
\begin{align}
F(\mathds{P}, \mathds{Q}) := \sum_{{\bf z}, {\bf z}', \errpe} \sqrt{\mathds{P}_{\bfZkey\bfZkeyp\Errpe}({\bf z}, {\bf z}', \errpe)\, \mathds{Q}_{\bfZkey\bfZkeyp\Errpe}({\bf z}, {\bf z}', \errpe)} = \sqrt{\Pr[\Errkey < 
  \Errpe + \mu\,|\textnormal{``pass''}]} \,,
\end{align}
  which can be bounded using~\eqref{eqn:bbkey}. 
  The purified distance between the distributions is then given by $P(\mathds{P}, \mathds{Q}) := \sqrt{1 - F^2(\mathds{P}, \mathds{Q})} = \eps'$.
  Hence, under the distribution $Q$, we have $\Errkey < \Errpe + \mu \leq \Qtol + \mu$ with certainty. 
  In particular, the total number of errors on $n$ bits, $W := n \Errkey$, satisfies
  \begin{align}
    \label{eqn:errorbound}
    W \leq \big\lfloor n (\Qtol + \mu) \big\rfloor \, .
  \end{align}
  The max-entropy, $\chmax{\bfZkey}{\bfZkeyp}{}$, is upper bounded by the minimum 
  number of bits of additional information about $\bfZkey$ needed to perfectly 
  reconstruct $\bfZkey$ from $\bfZkeyp$~\cite{renesrenner10}.
This value can in turn be upper bounded by the logarithm of the maximum support of $\bfZkey$ conditioned 
  on any value $\bfZkeyp$ = {\bf z}$'$. Since the total number of errors 
  under $Q$ satisfies~\eqref{eqn:errorbound}, we may write
  \begin{align}
    \chmaxeps{\bfZkey}{\bfZkeyp}{\mathds{P}} \leq
\chmax{\bfZkey}{\bfZkeyp}{\mathds{Q}}
      \leq \log \ \sum_{w = 0}^{\lfloor n (\Qtol + \mu) \rfloor} { n \choose w }
        \leq n h \big( \Qtol + \mu \big) \label{eqn:maxbound} \, .
  \end{align}
  The last inequality is shown in~\cite{vanlint99},
  Section 1.4. This
  concludes the proof of Lemma~\ref{lemma:stat}.
\end{proof}


\end{document}